\documentclass[5p,twocolumn,times,number]{elsarticle}

\usepackage{graphicx}
\usepackage{amsmath}  
\usepackage{hyperref} 
\usepackage{xcolor}
	
\def\pmbanner{{\hrule height 1 pt}\vskip35pt{NIMA POST-PROCESS BANNER TO BE REMOVED AFTER FINAL ACCEPTANCE}\vskip35pt{\hrule height 4pt}\vskip20pt}
\def\anue{$\bar{\nuup}_{\mathrm{e}}$}
\def\nue{${\nuup_\mathrm{e}}$}
\def\anumu{$\bar{\nuup}_{\muup}$}
\def\numu{$\nuup_{\muup}$}
\begin{document}

\begin{frontmatter}

\title{\pmbanner Neutrino-Antineutrino Identification in a Liquid Scintillator Detector: \\ Towards a Novel Decay-at-Rest-based Neutrino CPV Framework}

\author[add1]{M.~Grassi\corref{cor}}
\ead{marco.grassi@apc.in2p3.fr}
\author[add2]{F.~Pessina}
\author[add1,add3]{A.~Cabrera}
\author[add4]{S.~Dusini}
\author[add2]{H.~Nunokawa}
\author[add5,add1]{F.~Suekane}

\cortext[cor]{Corresponding author}

\address[add1]{AstroParticule et Cosmologie, CNRS/IN2P3, CEA/IRFU, Observatoire de Paris, Sorbonne
Paris Cit\'e, Paris, France}
\address[add2]{Pontificia Universidade Cat\'olica do Rio de Janeiro,
Departamento de F\'isica, Rio de Janeiro, Brazil}
\address[add3]{LNCA Underground Laboratory, IN2P3/CNRS - CEA, Chooz, France}
\address[add4]{INFN Sezione di Padova, Padova, Italy}
\address[add5]{Research Center for Neutrino Science, Tohoku University, Sendai, Japan}

\begin{abstract}
We introduce a novel approach to investigate CP violation in the neutrino sector, based on the simultaneous detection of \nue{} and \anue{} stemming from the oscillation of \numu{} and \anumu{} produced in the decay at rest of $\piup$s and $\muup$s at a beam target. This approach relies on a novel liquid scintillator detector technology expected to yield unprecedented identification of \nue{} and \anue{} charged-current interactions, which we investigate by means of Monte Carlo simulations. Here we report preliminary results concerning both the detection technique and its physics reach.
\end{abstract}


\end{frontmatter}

\section{Introduction}

One of the most compelling open questions in neutrino physics is the existence of CP violation in the leptonic sector, whose determination could be accessible through high-precision neutrino oscillation measurements. The most common experimental framework relies on measuring the difference between the appearance probability of \nue{} in a \numu{} beam, and the corresponding charge conjugate oscillation. 
Neutrino beams are typically in the GeV range and need to be operated in different configurations to produce \numu{} and \anumu{}, which implies large systematic uncertainties at the detection level~\cite{t2k}. 
A different approach is to rely on the production of \anumu{} only, and to detect the appearance of \anue{} at multiple baselines~\cite{dedalus}.

Here we introduce a CP violation framework based on the subsequent decay at rest (DAR) of $\piup$ and $\muup$ produced by colliding protons on a fixed target:
\begin{equation}
\begin{split}
\piup^+ \rightarrow \, \, & \muup^+ + \nuup_{\muup} \\
& \muup^+ \rightarrow \ \mathrm{e}^+ + \nuup_\mathrm{e} + \bar{\nuup}_{\muup}
\end{split}
\label{eq:dar}
\end{equation}
where, for the first time, both the $\nu_\muup \rightarrow \nu_e$ and the $\bar{\nu}_{\muup} \rightarrow \bar{\nu}_e$ channels are exploited \textit{simultaneously}.
This approach is geared to strongly reduce some of the most important systematic uncertainties affecting long-baseline neutrino experiments, and relies on a novel liquid scintillator (LS) detector technology expected to yield unprecedented identification of \nue{} and \anue{} charged-current interactions. This manuscript presents preliminary results ---based on simulations--- meant to assess both the capabilities of the new detection technique, and its physics reach.

\section{Neutrino Interaction and Detection}

The $\piup^+$ decay chain yields a 30 MeV \numu{}, and a \anumu{} with a continuous energy spectrum between 0 and 53 MeV.  At 30 MeV, the first oscillation maximum for the appearance of \anue{} and \nue{} occurs at a distance of 16 km, which we choose as a baseline for our simulated detector. 
When compared to long-baseline neutrino experiments, a DAR facility 
shows four main advantages: (1) the baseline is short enough to consider the $\nu$ oscillation happening in vacuum, meaning that uncertainties associated to the earth's density profile play a negligible role; (2) being produced as part of the same decay chain, the flux of  \numu{} and \anumu{} at the source is identical and isotropic, hence there is no uncertainty stemming from the contamination and the alignment of the $\nuup_\muup / \bar{\nuup}_\muup$ beams; 
(3) the energy of \anue{} and \nue{} is low enough to have the charged-current (CC) interactions being in the well-understood quasi-elastic regime; (4) the neutrino energy is below the production threshold of both $\muup$ and $\tauup$, therefore only \nue{} and \anue{} are detectable via charged-current interactions.

LS detectors have been used since the 50s to detect \anue{} through the Inverse Beta Decay (IBD) reaction ($\bar{\nuup}_\mathrm{e} + \mathrm{p} \rightarrow \mathrm{e}^+ + \mathrm{n}$), taking advantage of the large fraction of H present in organic LS~\cite{cowan}. The detection of \nue{}, on the contrary, is challenging because the \nue{} cross section on LS carbon is considerably lower than the IBD~\cite{scholberg}. To overcome this limitation we consider a  Pb-loaded LS, taking advantage of the $^{208}$Pb($\nuup_\mathrm{e}, \mathrm{e}^-$) CC interaction which, at 30 MeV, has a cross section 50 times larger than the IBD~\cite{scholberg}.

\begin{figure}
\centering
\includegraphics[width=\linewidth]{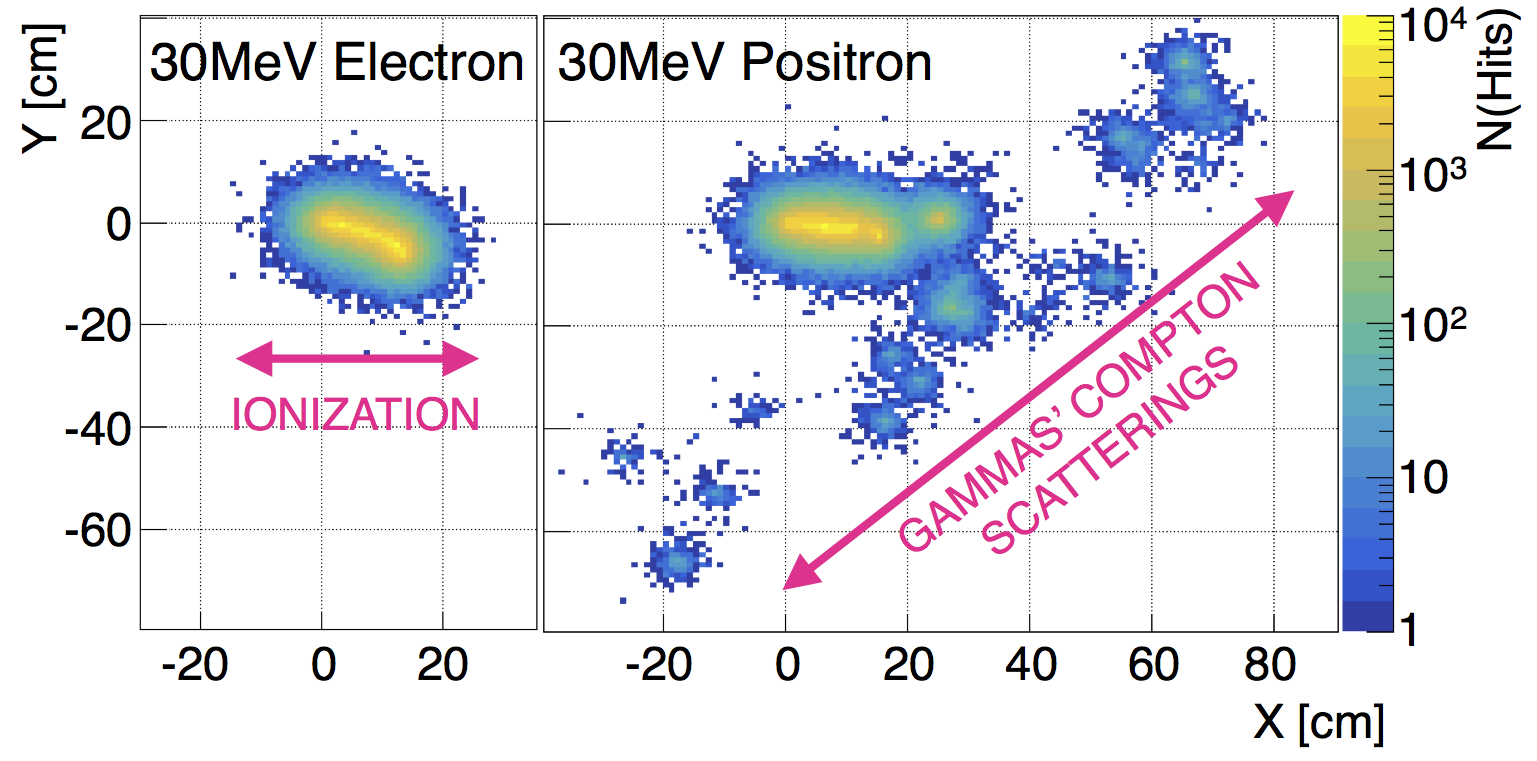}
\caption{Geant4 simulation of a single $\mathrm{e}^-$ (left) and $\mathrm{e}^+$ (right) event taking place in an opaque liquid scintillator detector, where the light is read out through a lattice of wavelength-shifting fibers arranged on the $xy$ plane and running along $z$. Each pixel is a fiber, and the color represents the number of scintillation photons impinging on it. Both the $\mathrm{e}^-$ and the $\mathrm{e}^+$ show a track-like energy deposition resulting from the particles' ionization. The multiple energy depositions due to Compton scattering of the positron's annihilation $\gammaup$s provide a unique signature allowing powerful $\mathrm{e}^{-}/\mathrm{e}^+$ tagging.}
\label{fig:ev_display}
\end{figure}

Tagging the charge of the final state lepton is pivotal to disentangle the \anue{} from the \nue{} sample. To this purpose,
we introduce a detector based on an opaque (translucent) LS~\cite{FJPPL}, where the scintillation light is not expected to travel up to the detector edges in order to be detected by photosensors. Rather, it gets collected close to the interaction vertex by a dense lattice of wavelength-shifting fibers crossing the whole detector. A full Geant4 simulation of single $\mathrm{e}^-$ (left) and $\mathrm{e}^+$ (right) events in the detector is shown in Fig.~\ref{fig:ev_display}. Each pixel in the figure represents a fiber, and the color pattern (z axis) indicates the number of scintillation photons impinging on a given fiber (hits). Thanks to a carefully tuned LS opacity, the scintillation light gets confined locally, and the energy depositions resulting from the Compton scattering of the $\mathrm{e}^+$'s annihilation $\gammaup$s are clearly distinguishable from the particle's $dE/dx$. We exploit such powerful $\gammaup$ identification to perform an efficient $\mathrm{e}^- / \mathrm{e}^+$ separation, while retaining a light level $\mathcal{O}$(200 photoelectrons/MeV).

The relative number of detected \nue{} and \anue{} depends primarily on the relative amount of Pb and H in the detector. Folding the cross section with the molar weight of the two elements we find that a 50\% Pb loading (by weight) results in the two rates of events to be identical. It is worth to stress that past experimental attempts to load metal compounds into LS by more than a few percent weight failed also because the resulting cocktail was loosing transparency, while in our case a reduced transparency would not compromise the light detection. Nevertheless, we report that our sensitivity to CP violation is preserved even with loading below 15\%, namely with the number of  \nue{} events being 30\% of the \anue{}.
The $\mathrm{e}^+/\mathrm{e}^-$ tagging efficiency could be compromised in the case of showering events, where the annihilation $\gammaup$s might be mimicked by bremsstrahlung radiation. An analysis to assess the reconstruction performance is ongoing. However, our simulation shows that the critical energy
$E_c$ of a 15\% Pb-loaded LS is around 50 MeV, implying that the majority of the detected events will have a deposited energy below $E_c$.

Radiogenic and cosmogenic backgrounds are uncorrelated with the beam spill. They can be heavily suppressed by applying a time selection based on the spill time, and can be further characterized and statistically subtracted using beam-off data. Beam-on backgrounds stem mostly from unoscillated \nue{} due to $\muup^+$ decay, and they are rejected using both their different timing and energy spectrum with respect to the signal.

\section{Results and Conclusion}
\vspace{-1mm}
We introduced a novel approach to look for the existence of CP violation in the neutrino sector based on exploiting for the first time both the \anumu{} and \numu{} produced in the decay at rest  of the $\piup^+$. This approach relies on a liquid scintillator detector which, for the first time, is able to distinguish charged-current \anue{} and \nue{} interactions without the help of a magnetic field. The detector, still in its R\&D phase, has been introduced through a full Geant4 simulation, showing how LS opacity can be tuned to perform powerful event imaging without compromising the calorimetry. The detector has the capability to be heavily loaded with metal compounds, hence allowing to boost \nue{} detection by means of interactions typically not present in organic LS. We investigated the capability of such a detection technique by means of a $\chi^2$ analysis, where both the rate and energy spectrum of the detected \nue{} and \anue{} events concur to the final sensitivity. We found that the main experimental systematic uncertainty stems from the poor knowledge of the dopant cross section (Pb in our case), and in our analysis we let its magnitude vary between 1\% and 10\%. We report that, using the current uncertainty on the mixing angle $\theta_{23}$~\cite{capozzi}, a 300 kton detector would be able to explore at the 5~$\sigma$ level 30\% to 60\% of the total $\delta_{CP}$ range in 10 years (depending on the assumed systematic uncertainty configuration).   

\vspace{-2mm}
\section*{Acknowledgments}

MG acknowledges support from the Marie Curie Research Grants Scheme, grant 707918.
FP is supported by CNPq Ph.D. scholarship.


\end{document}